\documentclass{svmult}
 \pdfoutput=1 

\usepackage[english]{babel}
\usepackage{makeidx} 
\usepackage{graphicx,wrapfig}
\usepackage{multicol}

\usepackage[valuesep=med,unitsep=thin]{siunitx}

\usepackage{caption}
\usepackage{rotating}

\usepackage[utf8]{inputenc}
\usepackage{amsmath}
\usepackage{amssymb}
\usepackage{colordvi}
\usepackage{color}
\usepackage{float}
\usepackage{graphicx}
\usepackage{ulem}
\usepackage{wrapfig} 
\usepackage{subcaption}
\usepackage{longtable,lscape}
\usepackage{multirow}
\usepackage{tabularx}

\usepackage{placeins}

\usepackage{ragged2e,array,longtable,lscape}
\usepackage{multirow}
\usepackage{tabularx}
\usepackage{booktabs}

\usepackage{ulem}

\usepackage{placeins}

\usepackage{ulem}
\usepackage{sidecap}

\usepackage{tikz} 
\usepackage{pgfplots}

\definecolor{DarkGreen}{rgb}{0.133333,0.545,0.1333}
\definecolor{DarkPurple}{rgb}{0.58 , 0.0 ,0.827}

\usepackage{ifthen}
\newboolean{createTikzExport}

\setboolean{createTikzExport}{false}
\setboolean{createTikzExport}{true}

\ifthenelse{\boolean{createTikzExport}}{
\usetikzlibrary{external}
\tikzset{external/system call={latex \tikzexternalcheckshellescape -halt-on-error -interaction=batchmode -jobname "\image" "\texsource";
dvips -o "\image".ps "\image".dvi;
ps2eps "\image.ps"}}
\tikzexternalize
}{}

\definecolor{DarkGreen}{rgb}{0.133333,0.545,0.1333}
\definecolor{DarkPurple}{rgb}{0.58 , 0.0 ,0.827}
\colorlet{darkgreen}{green!50!black}

\definecolor{maroon}{cmyk}{0,0.87,0.68,0.32}

\newcommand{\Nproc}{{\color{black}N_{\text{proc}}}} 
 
\newcommand{\Figure}{{\color{black}Fig.}}

\newcommand{\MPF}{w_k}
\newcommand{\PID}{\tau}
\def\twall{t_{\text{wall}}}
\def\tsim{t_{\text{sim}}}
\def\tmin{t_{\text{240}}}
\def\tNproc{t_{\Nproc}}
\newcommand{\figref}[1]{{\color{blue}Fig.}~\ref{#1}}

\def\E{{\mathbf E}}
\newcommand{\e}[1]{\cdot10^{#1}}

\def\x{{\mathbf x}}

\def\Ltot{{L_\text{tot}}}
\def\Lmean{{L_\text{mean}}}
\def\nProcs{{n_\text{Process}}}
\def\nElems{{n_\text{Elements}}}

\def\weight{{\nu}}
\def\np{{n_\text{Part}}}
\def\te{{t_\text{e}}}

\newcommand{\tm}[2]{{{t}_{m,\text{#1}}^{#2}}}

\newdimen\HilbertLastX
\newdimen\HilbertLastY
\newcounter{HilbertOrder}

\def\DrawToNext#1#2{%
   \advance \HilbertLastX by #1
   \advance \HilbertLastY by #2
   \pgfpathlineto{\pgfqpoint{\HilbertLastX}{\HilbertLastY}}
}

\def\Hilbert[#1,#2,#3,#4,#5,#6,#7,#8] {
  \ifnum\value{HilbertOrder} > 0%
     \addtocounter{HilbertOrder}{-1}
     \Hilbert[#5,#6,#7,#8,#1,#2,#3,#4]
     \DrawToNext {#1} {#2}
     \Hilbert[#1,#2,#3,#4,#5,#6,#7,#8]
     \DrawToNext {#5} {#6}
     \Hilbert[#1,#2,#3,#4,#5,#6,#7,#8]
     \DrawToNext {#3} {#4}
     \Hilbert[#7,#8,#5,#6,#3,#4,#1,#2]
     \addtocounter{HilbertOrder}{1}
  \fi
}

\def\hilbert((#1,#2),#3){%
   \advance \HilbertLastX by #1
   \advance \HilbertLastY by #2
   \pgfpathmoveto{\pgfqpoint{\HilbertLastX}{\HilbertLastY}}
     \setcounter{HilbertOrder}{#3}
     \Hilbert[10mm,0mm,-10mm,0mm,0mm,10mm,0mm,-10mm]
     \pgfusepath{stroke}%
  }

\normalsize
\renewcommand{\abstractname}{Abstract}
\begin{document}

\title*{
A Load Balance Strategy for Hybrid Particle-Mesh Methods
}
\titlerunning{
A load balance strategy for unstructured PIC simulations
 }
\author{
P.\,Ortwein\inst{1}\and
T.\,Binder\inst{2}\and
S.\,Copplestone\inst{1}\and
A.\,Mirza\inst{2}\and
P.\,Nizenkov\inst{2}\and
M.\,Pfeiffer\inst{2}\and
C.-D.\,Munz\inst{1}\and
S.\,Fasoulas\inst{2}
}
\authorrunning{Ortwein et al.}
\institute{
Institute of Aerodynamics and Gas Dynamics (IAG), University of Stuttgart, 70569 Stuttgart, Germany
\texttt{munz@iag.uni-stuttgart.de}
\and
Institute of Space Systems (IRS), University of Stuttgart, 70569 Stuttgart, Germany
\texttt{fasoulas@irs.uni-stuttgart.de}
}
\maketitle
\setcounter{footnote}{0}
\begin{abstract}
We present a load balancing strategy for hybrid particle-mesh methods that is based on domain 
decomposition and element-local time measurement. This new strategy is compared to our 
previous approach, which assumes a constant weighting factor for each particle to determine the 
computational load. The timer-based load balancing is applied to a plasma expansion simulation. The 
performance of the new algorithm is compared to results presented in the past and a significant 
improvement in terms of computational efficiency is shown.
\end{abstract}

\section{Introduction}\label{sec:introduction}
In highly parallelized simulation methods that are based on both particle and spatial 
discretization techniques, such as Particle-in-Cell (PIC) and Direct Simulation Monte Carlo (DSMC), 
load imbalance is inevitable. As soon as the particle distribution is inhomogeneous or certain types of boundary conditions are 
encountered, the load increases for certain regions within the computational domain. 
To avoid inefficient resource utilization, this problem has to be addressed, for example by using a 
flexible domain decomposition approach. 
In the following, the newest development of the PICLas code~\cite{Munz2014} concerning load 
balance concepts is presented. 
The approach is similar to the one presented in~\cite{Bettencourt2008}, where timers are used 
for determining a global ratio by calculating a particle-element weighting factor. 
A fixed value concept has therefore been implemented in previous versions of 
PICLas~\cite{ortwein2015}. 
However, a restriction of the redistribution scheme to a global particle-weighting factor is not 
suitable for all simulation scenarios. Therefore, individual weights on an element
level, which are determined dynamically during the simulation, are used in the present work. 
Other known load balance concepts for PIC migrate particles between MPI 
processes~\cite{Plimpton2003} or utilize a general particle-field decomposition 
approach~\cite{Qiang2010}. 
Most recently, a load-management strategy was proposed in~\cite{Beck2015}, including an improved 
data structure~\cite{Germaschewski2013} that enables an efficient distribution of the load, 
which is broken into individual segments. This approach is well suited for an efficient dynamic 
load distribution, similar to the element sorting along a space filling curve (SFC) in PICLas.

After a brief review of the underlying PIC theory in Sec.~\ref{sec:pic_theory}, the load computation 
and distribution concept is discussed in Sec.~\ref{sec:load_balance}. The results are presented in 
Sec.~\ref{sec:results} and the paper is concluded by a summary and outlook on future code developments 
in Sec.~\ref{sec:conclusions}.

\section{Particle-in-Cell Theory}\label{sec:pic_theory}
The well-known PIC method is typically used to find an approximate solution of the collision-free Boltzmann equation, which is also called the Vlasov equation:
\begin{equation}
  \dfrac{\partial f_{\alpha}}{\partial t}+\mathbf{v}_{\alpha}\dfrac{\partial f_{\alpha}}{\partial\mathbf{x}_{\alpha}}
  +\dfrac{\mathbf{F}}{m_{\alpha}}\dfrac{\partial f_{\alpha}}{\partial\mathbf{v}_{\alpha}}=0~.
  \label{eq:vlasoveq}
\end{equation}
Here, $f_{\alpha}=f_{\alpha}\left( \mathbf{x}, \mathbf{v},t \right)$ is the particle distribution function of species $\alpha$ at the position $\mathbf{x}$ and time $t$ 
 with velocity $\mathbf{v}$.
Additionally, $m$ is the particle mass, and $\mathbf{F}$ is the Lorentz force, given by
\begin{equation}
\mathbf{F}=q_{\alpha}\left(\mathbf{E}+\mathbf{v}_{\alpha}\times\mathbf{B}\right)~,
\end{equation}
with the particle charge $q$, the electric field $\mathbf{E}$ and the magnetic field $\mathbf{B}$.
The electromagnetic fields $\mathbf{E}$ and $\mathbf{B}$ are solutions of the Maxwell's equations 
\begin{align}
\frac{\partial \mathbf{E}}{\partial t} - c^2 \nabla \times \mathbf{B} & =  - \frac{\mathbf{j}}{\epsilon_0}~,\label{eq:max1} \\
\frac{\partial \mathbf{B}}{\partial t} + \phantom{c^2}\nabla \times \mathbf{E} & =  0~, \label{eq:max2}\\
\nabla \cdot \mathbf{E} & = \frac{\rho}{\epsilon_0}~,\label{eq:max3}\\
\nabla \cdot \mathbf{B} & =  0~. \label{eq:max4}
\end{align}
The corresponding source terms are the charge density $\rho$ and the current density $\mathbf{j}$, defined as moments of the distribution function by
\begin{align} 
  \begin{split}
  \rho(\mathbf{x},t) & = q \int_{\mathbb{R}^3}  f(\mathbf{x},\mathbf{v},t) \text d^3v~, \\
  \mathbf{j}(\mathbf{x},t) & = q  \int_{\mathbb{R}^3}  \mathbf{v}f(\mathbf{x},\mathbf{v},t) \text d^3v~. 
  \end{split} 
\label{eq:sourceterms}
\end{align}
The main idea of PIC methods is to approximate the distribution function by a discrete number of particles, which are mathematically described by 
the linear combination of $N$ $\delta$-functions with a super-particle weighting factor $\MPF$
\begin{equation}
f\left( \mathbf{x}, \mathbf{v},t \right)\approx \sum_{k=1}^N w_k \delta\left(\mathbf{x}-\mathbf{x}_k(t) \right)\delta\left(\mathbf{v}-\mathbf{v}_k(t) \right)~.
\label{eq:boltzapprox}
\end{equation}
Due to this approximation and the corresponding numerical errors, it is not sufficient to solely solve equations \eqref{eq:max1} and \eqref{eq:max2},
although they are mathematically well-defined for appropriate boundary and initial conditions. 
Owing to numerical errors, an additional unphysical divergence term appears. Consequently, 
equations \eqref{eq:max3} and \eqref{eq:max4} are no longer fulfilled. To overcome this problem, 
a purely hyperbolic formulation of Maxwell's equations (PHM) is solved as described in \cite{munz:2000-1}. The system of equations is given by
\begin{align}
  \frac{\partial \vec{ E}}{\partial t} &= c^2 \nabla \times \vec{ B}  - \frac{\vec{ j}}{\epsilon_0}-\chi c^2\nabla \Psi~,\label{eq:PHM1} \\
  \frac{\partial \vec{ B}}{\partial t} &= -\nabla \times \vec{ E} -\chi\nabla \Theta~, \label{eq:PHM2} \\
  \frac{\partial \Psi}{\partial t}  &=  \chi \left(-\nabla \cdot \vec{ E} +\frac{\rho}{\epsilon_0}\right)~, \label{eq:PHM3}\\
  \frac{\partial \Theta}{\partial t} &=  -\chi c^2 \nabla \cdot \vec{ B} \label{eq:PHM4}~,
\end{align}
with the dimensionless positive parameter $\chi$ and the generalized Lagrange multipliers 
$\Psi(\vec{x},t)$ and $\Theta(\vec{x},t)$, where the latter two properties are scalar potential 
fields. These additional variables couple the divergence conditions \eqref{eq:max3} and 
\eqref{eq:max4} to the evolution equations \eqref{eq:max1} and \eqref{eq:max2}.
The PHM system is solved using a Discontinuous Galerkin Spectral Element Method (DGSEM) as described in \cite{Munz2014}. 
In the following, a short introduction to this method is given. 
For the DG method, the simulation region $\Omega$ is split into non-overlapping grid cells 
$\Omega_i$ in which the approximation of the solution is a continuous function, 
usually a piecewise polynomial function. 
To solve equations~\eqref{eq:PHM1}-\eqref{eq:PHM4}, they are re-written 
in conservation form
\begin{equation}\label{eq:ConLaw}
\frac{\partial\textbf{U}}{\partial t}+\nabla\cdot\textbf{F}\left(\textbf{U}\right)=
\textbf{S}~,
\end{equation}
where $\textbf{F}$ represents the physical flux vector, $\textbf{U}$ the solution and $\textbf{S}$ the source terms, which are given by 
\begin{equation}
\textbf{U}=\left(\textbf{E},\textbf{B},\Psi,\Theta\right)^T~,
\end{equation}
and
\begin{equation}
  \textbf{S}=\frac{1}{\varepsilon_0}\left(\textbf{j},\vec{0},\chi\rho,0\right)^T~,
\end{equation} respectively.  In the DGSEM context, equation~\eqref{eq:ConLaw} is transformed from the physical space $\Omega$ to the reference 
space $\mathcal{E}\in[-1,1]^3$ giving
\begin{equation}
\frac{\partial\textbf{U}}{\partial t}+\frac{1}{J}\nabla_{\xi}\cdot\tilde{\textbf{F}}=\textbf{S}~,
\end{equation}
where $J$ is the Jacobian determinant of the transformation, $\nabla_{\xi}$ the divergence operator with respect to the 
reference space and $\tilde{\textbf{F}}$ the transformed flux vector. This equation is multiplied by a test function $\phi$ and 
integrated over $\mathcal{E}$ leading to
\begin{equation}\label{eq:DG}
\int_{\mathcal{E}}{J\frac{\partial\textbf{U}}{\partial t}~\phi}~d\boldsymbol\xi+
\underbrace{\int_{\partial \mathcal{E}}{\tilde{\textbf{F}}\cdot\textbf{N}~\phi}~d\textbf{s}}_{\text{Surface Integral}}
-
\underbrace{\int_{\mathcal{E}}{\tilde{\textbf{F}}\cdot\nabla_{\xi}\phi}~d\boldsymbol\xi}_{\text{Volume Integral}}
=
\int_{\mathcal{E}}{J\textbf{S}~\phi}~d\boldsymbol\xi~,
\end{equation}
where integration by parts has been used to split the divergence integral into a surface and a volume integral. 
The advantage of this method in the context of parallelization is that only the surface integral is responsible for the inter-cell
coupling between DGSEM cells.
Using MPI parallelization, the only messages that have to be communicated arise from this 
surface integral. This leads to a highly efficient scheme with remarkable scaling properties in 
high performance computing (HPC) contexts \cite{hindenlang:2013}.

\section{Load Balance Strategy}\label{sec:load_balance}
The present load measurement strategy is applicable for two reasons: Firstly, the high-order DG 
method as used and, secondly, an explicit time-stepping. As the use of high-order discretization tends to enlarge elements, a 
particle remains in a certain element for a longer period of time. This justifies the utilization of
a previously determined distribution over a certain simulation period. In the following, the load computation
and distribution scheme is presented in detail.
\subsection{Load Computation}
\label{sec:load_computation}
In order to distribute the load between different processes, the total load $\Ltot$ has to be 
determined. It is computed by adding up the load $L_{i}$ that is ascertained for each individual 
element 
\begin{equation}
  \Ltot = \sum_{i=1}^{\nElems} L_{i}~.
\end{equation}
In a perfectly balanced simulation, each MPI process receives the average load
\begin{equation}
  L_{average} = \frac{\Ltot}{\nProcs}~,
\end{equation}
with $\nProcs$ being the number of MPI processes.
The load per element $L_{i}$ can be determined by different methods: particle weighting or runtime 
measurements. The first idea uses a fixed weight $\weight$ for each particle. Hence, the load of each 
cell is calculated by 
\begin{equation}
L_{i} = 1 + \weight~\np,
\end{equation}
where the assumed constant load of the DG operator is increased by $\weight~\np$, which is the particle-element
weight multiplied by the number of particles that reside in each cell. The latter is  
run-time dependent and represents merely a rule of thumb. Fixed particle weights are simple to 
implement. However, they have to be determined for each simulation scenario. 
Amongst others, they depend on the number of simulated particles and boundary interactions. 
Runtime measurements, on the other hand, are more flexible, allowing to compute the particle-element weight 
during the simulation, which inherently considers inhomogeneous particle distributions and varying 
computational effort within the domain. The measurement utilized here exclusively contains 
computational time; hence, MPI communication is not considered. 
MPI communication and process-idle time depend on the current load distribution and are 
neglected, since an ideal non-blocking communication should introduce no overheads. To account for 
region-dependent loads, runtime measurement is not performed at process level, but rather the 
computational time for each element $\te$ is measured. 
In this work, the total computation time for each element consists of different components for 
each module and is defined as
\begin{equation}
\label{eq:elemtime}
\te = \tm{Field}{}+\tm{Particle}{}~,
\end{equation}
with $\tm{Field}{}$ being the time for the field solver and $\tm{Particle}{}$ the time for the particle treatment.
For $\tm{Field}{}$, the time is measured in total for each process and the process average time is assigned to each 
element, assuming the required computational effort is similar for each cell. In contrast, $\tm{Particle}{}$ depends
on the considered element. The time measurement directly provides the load distribution throughout the 
computational domain. In a next step, the load of each element is set equal to the measured time
\begin{equation}
\label{eq:load-per-elemtime}
  L_{i} = \te~,
\end{equation}
and the load is distributed as described in the following section.
After a specific time interval or number of iterations, the current load is compared with the 
last measured load and if the deviation
\begin{equation}
\label{eq:load-deviation}
  \Delta L = \cfrac{L_{tot,old}-L_{tot,new}}{L_{tot,old}}>\alpha~,
\end{equation}
is above a threshold $\alpha$, a load distribution step is performed. 
\subsection{Load Distribution}
\label{sec:load_distribution}
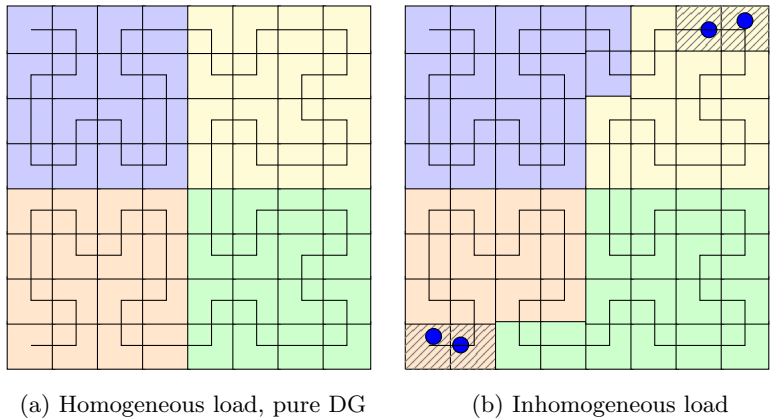
\begin{figure}
  \def\SizeW{0.44\textwidth}
  \begin{subfigure}[b]{0.44\textwidth}
    \centering
    \begin{tikzpicture}[scale=0.6,every node/.style={transform shape}]
 \foreach \i in {1,...,16}
  {
      \pgfmathtruncatemacro{\y}{(\i-1 ) / 4};
      \pgfmathtruncatemacro{\x}{\i -1 - 4 * \y};
      \pgfmathtruncatemacro{\label}{\x +7 * (7 - \y)};
      \node[rectangle,draw=black,fill=white!80!orange,minimum size=30] () at (\x,\y) {};
  }
  \foreach \i in {17,...,32}
  {
      \pgfmathtruncatemacro{\y}{(\i-1 ) / 4};
      \pgfmathtruncatemacro{\x}{\i -1 - 4 * \y};
      \pgfmathtruncatemacro{\label}{\x +7 * (7 - \y)};
      \node[rectangle,draw=black,fill=white!80!blue,minimum size=30] () at (\x,\y) {};
  }
  \foreach \i in {1,...,16}
  {
      \pgfmathtruncatemacro{\y}{(\i-1 ) / 4};
      \pgfmathtruncatemacro{\x}{\i +3 - 4 * \y};
      \pgfmathtruncatemacro{\label}{\x +7 * (7 - \y)};
      \node[rectangle,draw=black,fill=white!80!green,minimum size=30] () at (\x,\y) {};
  }
  \foreach \i in {17,...,32}
  {
      \pgfmathtruncatemacro{\y}{(\i-1 ) / 4};
      \pgfmathtruncatemacro{\x}{\i +3 - 4 * \y};
      \pgfmathtruncatemacro{\label}{\x +7 * (7 - \y)};
      \node[rectangle,draw=black,fill=white!80!yellow,minimum size=30] () at (\x,\y) {};
  }
 \hilbert((0mm,0mm),3)
\end{tikzpicture}
    \caption{Homogeneous load, pure DG}
    \label{fig:hilbert-domain}
  \end{subfigure}
  \def\SizeW{0.44\textwidth}
  \begin{subfigure}[b]{0.44\textwidth}
    \usetikzlibrary{patterns}
\begin{tikzpicture}[scale=0.6,every node/.style={transform shape}]
  \foreach \i in {1,...,2}
   {
       \pgfmathtruncatemacro{\y}{(\i-1 ) / 4};
       \pgfmathtruncatemacro{\x}{\i -1 - 4 * \y};
       \pgfmathtruncatemacro{\label}{\x +7 * (7 - \y)};
       \node[rectangle,draw=black,fill=white!80!orange,minimum size=30] () at (\x,\y) {};
       \node[pattern=north east lines,pattern color=gray,minimum size=30] () at (\x,\y) {};
   }
  \foreach \i in {5,...,16}
   {
       \pgfmathtruncatemacro{\y}{(\i-1 ) / 4};
       \pgfmathtruncatemacro{\x}{\i -1 - 4 * \y};
       \pgfmathtruncatemacro{\label}{\x +7 * (7 - \y)};
       \node[rectangle,draw=black,fill=white!80!orange,minimum size=30] () at (\x,\y) {};
   }

  \foreach \i in {3,...,4}
   {
       \pgfmathtruncatemacro{\y}{(\i-1 ) / 4};
       \pgfmathtruncatemacro{\x}{\i -1 - 4 * \y};
       \pgfmathtruncatemacro{\label}{\x +7 * (7 - \y)};
       \node[rectangle,draw=black,fill=white!80!green,minimum size=30] () at (\x,\y) {};
   }

   \foreach \i in {17,...,32}
   {
       \pgfmathtruncatemacro{\y}{(\i-1 ) / 4};
       \pgfmathtruncatemacro{\x}{\i -1 - 4 * \y};
       \pgfmathtruncatemacro{\label}{\x +7 * (7 - \y)};
       \node[rectangle,draw=black,fill=white!80!blue,minimum size=30] () at (\x,\y) {};
   }
   \foreach \i in {29,...,29}
   {
       \pgfmathtruncatemacro{\y}{(\i-1 ) / 4};
       \pgfmathtruncatemacro{\x}{\i +3 - 4 * \y};
       \pgfmathtruncatemacro{\label}{\x +7 * (7 - \y)};
       \node[rectangle,draw=black,fill=white!80!blue,minimum size=30] () at (\x,\y) {};
   }
   \foreach \i in {25,...,25}
   {
       \pgfmathtruncatemacro{\y}{(\i-1 ) / 4};
       \pgfmathtruncatemacro{\x}{\i +3 - 4 * \y};
       \pgfmathtruncatemacro{\label}{\x +7 * (7 - \y)};
       \node[rectangle,draw=black,fill=white!80!blue,minimum size=30] () at (\x,\y) {};
   }

   \foreach \i in {1,...,16}
   {
       \pgfmathtruncatemacro{\y}{(\i-1 ) / 4};
       \pgfmathtruncatemacro{\x}{\i +3 - 4 * \y};
       \pgfmathtruncatemacro{\label}{\x +7 * (7 - \y)};
       \node[rectangle,draw=black,fill=white!80!green,minimum size=30] () at (\x,\y) {};
   }
   \foreach \i in {30,...,32}
   {
       \pgfmathtruncatemacro{\y}{(\i-1 ) / 4};
       \pgfmathtruncatemacro{\x}{\i +3 - 4 * \y};
       \pgfmathtruncatemacro{\label}{\x +7 * (7 - \y)};
       \node[rectangle,draw=black,fill=white!80!yellow,minimum size=30] () at (\x,\y) {};
   }
   \foreach \i in {31,...,32}
   {
       \pgfmathtruncatemacro{\y}{(\i-1 ) / 4};
       \pgfmathtruncatemacro{\x}{\i +3 - 4 * \y};
       \pgfmathtruncatemacro{\label}{\x +7 * (7 - \y)};
       \node[pattern=north east lines,pattern color=gray,minimum size=30] () at (\x,\y) {};
   }
   \foreach \i in {17,...,24}
   {
       \pgfmathtruncatemacro{\y}{(\i-1 ) / 4};
       \pgfmathtruncatemacro{\x}{\i +3 - 4 * \y};
       \pgfmathtruncatemacro{\label}{\x +7 * (7 - \y)};
       \node[rectangle,draw=black,fill=white!80!yellow,minimum size=30] () at (\x,\y) {};
   }
   \foreach \i in {26,...,28}
   {
       \pgfmathtruncatemacro{\y}{(\i-1 ) / 4};
       \pgfmathtruncatemacro{\x}{\i +3 - 4 * \y};
       \pgfmathtruncatemacro{\label}{\x +7 * (7 - \y)};
       \node[rectangle,draw=black,fill=white!80!yellow,minimum size=30] () at (\x,\y) {};
   }

  \node[circle,draw=black,fill=blue,minimum size=10] (p1) at (0.1,0.2) {};
  \node[circle,draw=black,fill=blue,minimum size=10] (p2) at (0.7,0.01) {};
  \node[circle,draw=black,fill=blue,minimum size=10] (p3) at (6.2,7.0) {};
  \node[circle,draw=black,fill=blue,minimum size=10] (p4) at (7.,7.2) {};
  \hilbert((0mm,0mm),3)
\end{tikzpicture}
    \caption{Inhomogeneous load}
    \label{fig:hilbert-particle}
  \end{subfigure}
  \caption{Domain decomposition via SFC ordering.}
   \label{fig:domain-all}
\end{figure}

The load distribution between the MPI processes is a crucial step. If the load is not distributed homogeneously, 
the load balancing effect is limited. 
In a preprocessing step, all elements within the computational domain are sorted along a Hilbert 
curve due to its clustering property \cite{moon200101}. Then, each MPI process receives a certain 
segment of the SFC. 
To illustrate an optimal load balance scenario, a simplified grid is considered that consists of 
$8 \times 8 = 64$ elements, which are ordered along a SFC. \Figure~\ref{fig:hilbert-domain} depicts 
the decomposition of the grid into four regions, each corresponding to an individual MPI 
process. Without particles, no load imbalance is observed.
Next, in the lower left and upper right corner, four particles are inserted thus increasing the 
total computational weight from $64$ to $72$ (arbitrary units) using a particle-element weight of 2. A domain decomposition algorithm 
is then applied and the elements and its particles are assigned to a certain process.
To maintain load balance, the size of two empty regions has to increase, whereas the other two have 
to decrease, which is achieved by moving the element assignment along the SFC to the next MPI 
process. The results of this balancing step are shown in \Figure~\ref{fig:hilbert-particle},
and the total domain still remains balanced. However, the introduction of the load imbalance at 
the element level yields additional difficulties. The first case can be distributed among 
$\{2,4,8,16,32,64\}$ MPI processes, whereas the latter, imbalanced case can only be distributed on 
$\{2,4,8\}$ processes in order to be fully balanced. This example demonstrates different aspects. 
The load has to be assigned carefully, possible load imbalances reduce the total number of 
applicable MPI processes and therefore the load balance may not be guaranteed for all cases.
From these considerations, specifications for the load distribution algorithm can be defined, 
e.g., it has to satisfy the following two conditions
\begin{enumerate}
  \item Minimize the load deviation,
  \item Prevent the last process from receiving to much load (above average).
\end{enumerate}
By applying the first condition, each process $p$ receives a load 
\begin{equation}
  \label{eq:load-distribution}
  L_p=\bigg| L_\mathrm{target} - \sum_{i_\text{low}}^{i_\text{up}} L_i \bigg| \overset{!}{=} \min,
\end{equation}
where $i_\text{low}$ is the lower element index that is not assigned previously to a certain 
MPI process. Here, $i_\text{up}$ is the upper element index assigned to process $p$ that is 
chosen to minimize the deviation from the average load for this process. 
Next, it is examined if the last MPI process receives a load larger than the average value. 
If this condition is satisfied, then the mean deviation is increased by the load difference 
\begin{equation}
  \label{eq:load-adapt}
  L_\mathrm{target}^{0} = \Lmean,\qquad L_\mathrm{target}^{i+1} = L_\mathrm{target}^{i} + \frac{L_\text{last}-L_\mathrm{target}^{i}}{\nProcs},
\end{equation}
and the load distribution step is repeated. The second condition guarantees that the last 
process receives a smaller value than the average load. Thus, not acquiring too much load prevents the introduction   
of a load imbalance by the distribution algorithm. In an optimally balanced case, each process receives 
approximately the average load. Due to the sequential load distribution, it cannot be guaranteed that the last 
process receives the average load. Hence, it must be prevented that the last process receives a load 
larger than the average in order to circumvent idle time.

\section{Results}\label{sec:results}
\begin{figure}[!h]
\sidecaption
\includegraphics[trim = 0mm 0mm 0mm 0mm, clip, width=6cm]{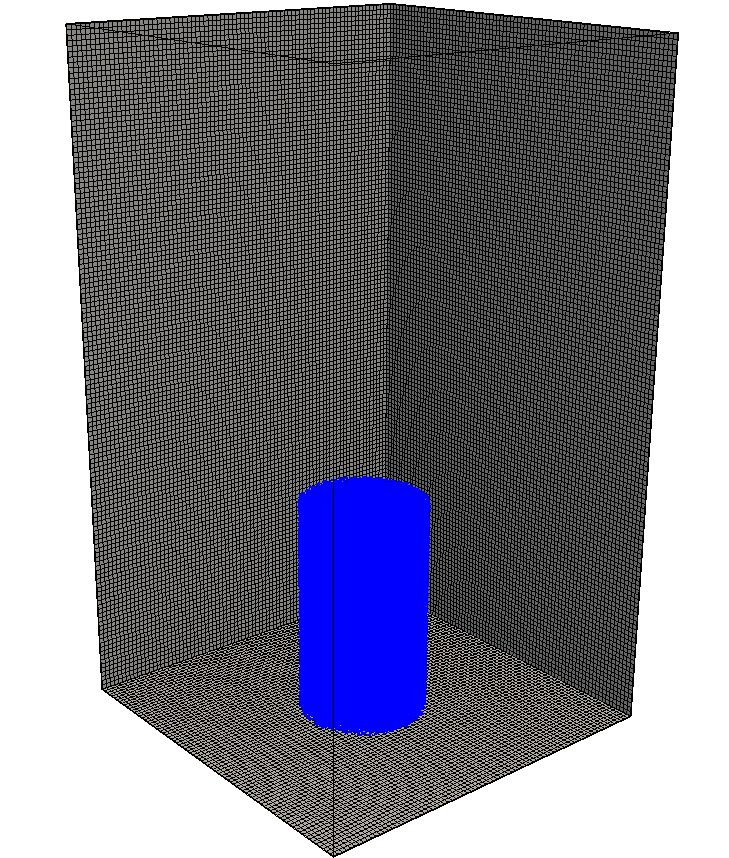}
\caption{Initial particle distribution and cuboid grid utilizing 80$\times$80$\times$140 cells.}
\label{fig_plume_initial_condition}
\end{figure}
A plasma plume setup from~\cite{Copplestone2016} is revised to investigate the improvements of the new load 
balancing concept. The setup consists of a cylindrical plasma (radius $r=20\mu $m, height 
$h_z=70\mu $m in $z$-direction) that is placed inside a cuboid domain of
$\Delta x\times\Delta y\times\Delta z=120\mu\text{m}\times120\mu\text{m}\times210\mu\text{m}$. 
This results in a heterogeneous particle distribution and consequently in a heterogeneously distributed load.
The presented results consider a small simulation time $\tsim$ within which the particles 
movement does not lead to a significant change in spatial load distribution, thus, the static 
load balance limit is investigated. Additionally, the cases are pure PIC simulations without DSMC routines.
It is examined whether the load can be distributed for a given number of cells and 
particles among the chosen number of MPI processes, where each process is always assigned to a single core. The 
initial conditions of both setups are given in Table~\ref{tab:settings}.
\begin{table}[!h]
\renewcommand{\tabcolsep}{0.4cm}
  \caption{Initial Conditions}
  \begin{center}
    \begin{tabular}{lcc}
      \toprule
      Property & Case A & Case B \\
      \midrule
      Particle number      & $N_\mathrm{e^-}=N_\mathrm{Al^+}=1\e{5}$ & $N_\mathrm{e^-}=N_\mathrm{Al^+}=1\e{6}$\\
      Electron density     & $n_\mathrm{e^-}=5\e{24}$ m$^{-3}$ & $n_\mathrm{e^-}=3.41\e{23}$ m$^{-3}$\\ 
      Simulation particle weight      & $\MPF\approx4.4\e{6}$ & $\MPF\approx3.0\e{4}$\\ 
      Electron temperature & $T_\mathrm{e^-}=2\e{4}$ K & $T_\mathrm{e^-}=1.16\e{7}$ K \\ 
      Ion temperature      & $T_\mathrm{Al^+}=1\e{4}$ K & $T_\mathrm{Al^+}=1.16\e{5}$ K\\ 
      Debye length         & $\lambda_D=2.52\e{-9}$ m & $\lambda_D=4.0\e{-7}$ m\\
      Polynomial degree     & 6                       & 3 \\
      \bottomrule
    \end{tabular}
  \end{center}
  \label{tab:settings} 
\end{table}
%
%
The algorithms parallel performance is measured by 
\begin{equation}
\PID=\frac{\tsim}{\Nproc\twall}~,
\end{equation}
where $\tsim$ is the simulated time difference, $\twall$ the wall clock time (without I/O and 
initialization) and $\Nproc$ the number of processes utilized within the simulation. 
It relates the simulation time to the employed resources, which are the total number of CPU hours used for a simulation. 
\figref{fig_plume_performance} illustrates the results obtained for two different deposition 
methods, a delta distribution and a shape function and compares them to the previous code 
version~\cite{Copplestone2016}. 
For the $\delta$-function deposition, the new code performs similarly to the old version. However, 
when a shape-function is utilized the performance increases. In this example, the new code outperforms the 
previous version by a factor of three, presumably, due to latency hiding.
For $\Nproc\leq4800$, the performance is kept at a constant level and 
when $\Nproc>4800$, the problem size (memory requirement) for each MPI domain is decreased for 
which caching effects lead to an even better performance.
As soon as the one cell per MPI domain limit is reached, the work within this cell cannot 
be parallelized in the current framework and currently represents the algorithm's barrier.
\def\TheRef{[3]}
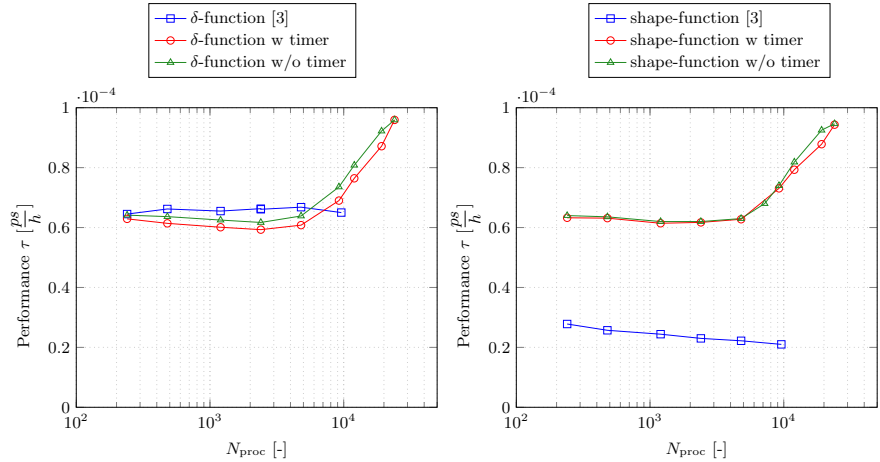
\begin{figure}[!h]
\sidecaption
\begin{tikzpicture}[scale=0.7]
\begin{semilogxaxis}[
font=\footnotesize,
grid = minor,
grid=both,
grid style={dotted},
xmin = 100,
xmax = 50000,
ymin = 0,
ymax = 10e-5,
xlabel={$\Nproc$~[-]},
ylabel={Performance $\PID$~[$\dfrac{ps}{h}$]},
ylabel style={at={(0.04,0.7)}, anchor=east},
legend cell align=left,
legend pos= north east,
legend style={at={(0.2,1.1)},anchor=south west}
] 
\addplot[mark=square, color=blue] table[header=false,x expr=\thisrowno{0},x index=0,y index=11,col sep=comma]
{./data/plume_scaling_2015_delta.csv};
\addlegendentry{$\delta$-function~\TheRef}
\addplot[mark=o, color=red] table[header=false,x expr=\thisrowno{1},x index=1,y index=11,col sep=comma]
{./data/plume_scaling_2016_delta_with_LB.csv};
\addlegendentry{$\delta$-function w timer}
\addplot[mark=triangle, color=DarkGreen] table[header=false,x expr=\thisrowno{1},x index=1,y index=11,col sep=comma]
{./data/plume_scaling_2016_delta_without_LB.csv};
\addlegendentry{$\delta$-function w/o timer}
\end{semilogxaxis}
\end{tikzpicture}
\begin{tikzpicture}[scale=0.7]
\begin{semilogxaxis}[
font=\footnotesize,
grid = minor,
grid=both,
grid style={dotted},
xmin = 100,
xmax = 50000,
ymin = 0,
ymax = 10e-5,
xlabel={$\Nproc$~[-]},
ylabel={Performance $\PID$~[$\dfrac{ps}{h}$]},
ylabel style={at={(0.04,0.7)}, anchor=east},
legend cell align=left,
legend pos= north east,
legend style={at={(0.2,1.1)},anchor=south west}
] 
\addplot[mark=square, color=blue] table[header=false,x expr=\thisrowno{0},x index=0,y index=11,col sep=comma]
{./data/plume_scaling_2015_shape.csv};
\addlegendentry{shape-function~\TheRef}
\addplot[mark=o, color=red] table[header=false,x index=1,y index=11,col sep=comma]
{./data/plume_scaling_2016_shape_with_LB.csv};
\addlegendentry{shape-function w timer}
\addplot[mark=triangle, color=DarkGreen] table[header=false,x expr=\thisrowno{1},x index=1,y index=11,col sep=comma]
{./data/plume_scaling_2016_shape_without_LB.csv};
\addlegendentry{shape-function w/o timer}
\end{semilogxaxis}
\end{tikzpicture}
\caption{Code peak performance for two deposition techniques as compared with~\cite{Copplestone2016}, where the number 
of processes chosen were between $240$ and $24000$ (Case A).}
 \label{fig_plume_performance}
\end{figure}
\def\TheRef{[3]}
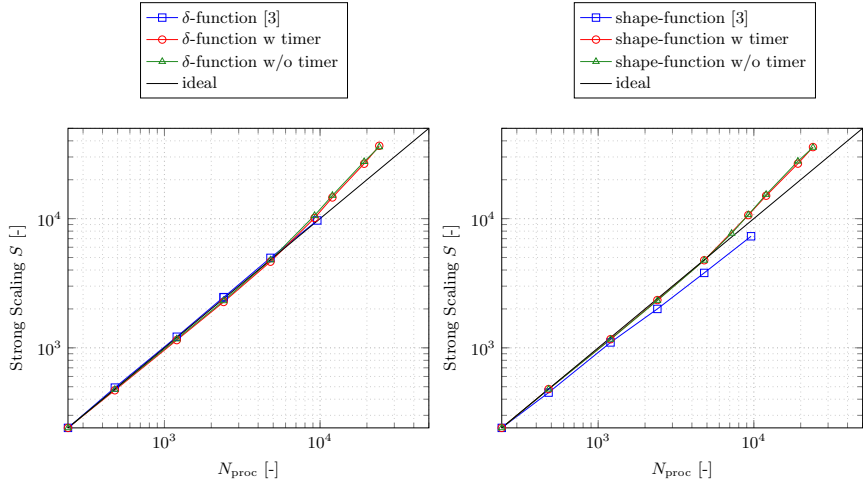
\begin{figure}[!h]
\sidecaption
\begin{tikzpicture}[scale=0.7]
\begin{loglogaxis}[
font=\footnotesize,
grid = minor,
grid=both,
grid style={dotted},
xmin = 240,
xmax = 50000,
ymin = 240,
ymax = 50000,
xlabel={$\Nproc$~[-]},
ylabel={Strong Scaling $S$~[-]},
ylabel style={at={(0.04,0.7)}, anchor=east},
legend cell align=left,
legend pos= north east,
legend style={at={(0.2,1.1)},anchor=south west}
] 
\addplot[mark=square, color=blue] table[header=false,x expr=\thisrowno{0},x index=0,y index=1,col sep=comma]
{./data/plume_scaling_2015_delta.csv};
\addlegendentry{$\delta$-function~\TheRef}
\addplot[mark=o, color=red] table[header=false,x expr=\thisrowno{1},x index=1,y index=0,col sep=comma]
{./data/plume_scaling_2016_delta_with_LB.csv};
\addlegendentry{$\delta$-function w timer}
\addplot[mark=triangle, color=DarkGreen] table[header=false,x expr=\thisrowno{1},x index=1,y index=0,col sep=comma]
{./data/plume_scaling_2016_delta_without_LB.csv};
\addlegendentry{$\delta$-function w/o timer}
\addplot[mark=none,domain=1:50000] {x};
\addlegendentry{ideal}
\end{loglogaxis}
\end{tikzpicture}
\begin{tikzpicture}[scale=0.7]
\begin{loglogaxis}[
font=\footnotesize,
grid = minor,
grid=both,
grid style={dotted},
xmin = 240,
xmax = 50000,
ymin = 240,
ymax = 50000,
xlabel={$\Nproc$~[-]},
ylabel={Strong Scaling $S$~[-]},
ylabel style={at={(0.04,0.7)}, anchor=east},
legend cell align=left,
legend pos= north east,
legend style={at={(0.2,1.1)},anchor=south west}
] 
\addplot[mark=square, color=blue] table[header=false,x expr=\thisrowno{0},x index=0,y index=1,col sep=comma]
{./data/plume_scaling_2015_shape.csv};
\addlegendentry{shape-function~\TheRef}
\addplot[mark=o, color=red] table[header=false,x index=1,y index=0,col sep=comma]
{./data/plume_scaling_2016_shape_with_LB.csv};
\addlegendentry{shape-function w timer}
\addplot[mark=triangle, color=DarkGreen] table[header=false,x expr=\thisrowno{1},x index=1,y index=0,col sep=comma]
{./data/plume_scaling_2016_shape_without_LB.csv};
\addlegendentry{shape-function w/o timer}
\addplot[mark=none,domain=1:50000] {x};
\addlegendentry{ideal}
\end{loglogaxis}
\end{tikzpicture}
\caption{Optimum strong scaling for two deposition techniques as compared with~\cite{Copplestone2016}, where the number 
of processes chosen were between $240$ and $24000$ (Case A).}
 \label{fig_plume_scaling}
\end{figure}
%
\figref{fig_plume_scaling} depicts the parallel strong scaling of the code up to $24,000$ cores 
for the new and old version of the code, which relates the speed-up due to parallelization 
\begin{equation}
S=\frac{240 \cdot \tmin}{\tNproc}~,
\end{equation}
where $\tmin$ is the wall time for a simulation with a minimum number of 10 compute nodes, each 
yielding 24 physical cores, as they offer the minimum amount of memory required for the test 
case and $\tNproc$ is the simulation time for a setup with $\Nproc$ processes.
\input{./graphs/fig_plume_performance_p3.tex}
For the next setup, the polynomial degree is reduced to $N=3$ and the number of particles 
per species is increased by a factor of ten to increase the computational load of the particles. 
The left-hand diagram of \figref{fig_plume_performance_p3} depicts a slice in the $x-z$-plane and 
shows the computational time per DG element. The inner region is particle-laden, and each element 
is roughly one hundred times more expensive than a pure DG element. 
Additionally, the computational time shows variation for pure DG elements.
The right-hand graph in \figref{fig_plume_performance_p3} illustrates the performance with and without 
the element time measurement. A fixed particle-element weight ($\weight=$ 0.02) results in a constant performance 
distribution over the investigated range of MPI processes (240 to 9216). 
In contrast, the performance decreases with higher number of MPI processes when a timer is 
utilized. Nevertheless, the performance is more than three to five times higher than without time 
measurement.
\begin{figure}[!h]
\def\MS{1}
\sidecaption
\begin{tikzpicture}[scale=0.65]
\begin{axis}[
  axis y line*=left,
font=\footnotesize,
grid = minor,
grid=both,
grid style={dotted},
xmin = -10,
xmax = 260,
ymin = 0,
ymax = 1.25,
xlabel={rank~[-]},
ylabel={$L_{\text{proc}}/L_{\text{mean}}$~[-]},
ylabel style={at={(0.04,0.7)}, anchor=east},
legend cell align=left,
legend pos= north east,
legend style={red,at={(0.2,1.1)},anchor=south west},
        yticklabel style=red,
        ylabel style=red,
        y axis line style=red,
        ytick style=red
] 
\addplot[mark=+, mark size=\MS, color=red,only marks] table[header=true,x expr=\thisrowno{0},x index=0,y expr=\thisrowno{3}/179.71,y index=3,col sep=space]{./data/p6_partitionInfo-240.txt};
\end{axis}
\begin{semilogyaxis}[
  axis y line*=right,
  axis x line=none,
  xmin = -10,
  xmax = 260,
  ymin=1e-5, ymax=1e2,
  ylabel style={blue,at={(1.35,0.15)}, anchor=west},
        yticklabel style=blue,
        ylabel style=blue,
        y axis line style=blue,
        ytick style=blue
]
\addplot[mark=+, mark size=\MS, color=blue,only marks] table[header=true,x expr=\thisrowno{0},x index=0,y expr=(\thisrowno{3}-179.7)/179.71,y index=3,col sep=space]{./data/p6_partitionInfo-240.txt};
\end{semilogyaxis}
\end{tikzpicture}
\begin{tikzpicture}[scale=0.65]
\begin{axis}[
  axis y line*=left,
font=\footnotesize,
grid = minor,
grid=both,
grid style={dotted},
xmin = -1000,
xmax = 26000,
ymin = 0,
ymax = 1.25,
xlabel={rank~[-]},
ylabel style={at={(0.04,0.7)}, anchor=east},
legend cell align=left,
legend pos= north east,
legend style={red,at={(0.2,1.1)},anchor=south west},
        yticklabel style=red,
        ylabel style=red,
        y axis line style=red,
        ytick style=red
] 
\addplot[mark=+, mark size=\MS, color=red,only marks] table[header=false,x expr=\thisrowno{0},x index=0,y expr=\thisrowno{1}/37.50,y index=1,col sep=space]{./data/p6_partitionInfo-24000_38_n10.csv};
\addplot[mark=+, mark size=\MS, color=red,only marks] table[header=false,x expr=\thisrowno{0},x index=0,y expr=\thisrowno{1}/37.50,y index=1,col sep=space]{./data/p6_partitionInfo-24000_removed_cols.csv};
\end{axis}
\begin{semilogyaxis}[
  axis y line*=right,
  axis x line=none,
  xmin = -1000,
  xmax = 26000,
  ymin=1e-5, ymax=1e2,
  ylabel={$(L_{\text{proc}}-L_{\text{median}})/L_{\text{mean}}$~[-]},
  ylabel style={blue,at={(1.35,0.15)}, anchor=west},
        yticklabel style=blue,
        ylabel style=blue,
        y axis line style=blue,
        ytick style=blue
]
\addplot[mark=+, mark size=\MS, color=blue,only marks] table[header=false,x expr=\thisrowno{0},x index=0,y expr=(\thisrowno{1}-38)/37.50,y index=1,col sep=space]{./data/p6_partitionInfo-24000_38_n10.csv};
\addplot[mark=+, mark size=\MS, color=blue,only marks] table[header=true,x expr=\thisrowno{0},x index=0,y expr=(\thisrowno{1}-38)/37.50,y index=1,col sep=space]{./data/p6_partitionInfo-24000_removed_cols.csv};
\end{semilogyaxis}
\end{tikzpicture}
\caption{Relative measured load (Case A) over MPI ranks (red) and positive deviation from median in 
logarithmic scale (blue), and the number of processes are $240$ (left) and $24000$ (right).}
\label{fig_plume_pload_distribution_p6}
\end{figure}
\begin{figure}[!h]
\def\MS{1}
\sidecaption
\begin{tikzpicture}[scale=0.65]
\begin{axis}[
  axis y line*=left,
font=\footnotesize,
grid = minor,
grid=both,
grid style={dotted},
xmin = -10,
xmax = 260,
ymin = 0,
ymax = 1.25,
xlabel={rank~[-]},
ylabel={$L_{\text{proc}}/L_{\text{mean}}$~[-]},
ylabel style={at={(0.04,0.7)}, anchor=east},
legend cell align=left,
legend pos= north east,
legend style={red,at={(0.2,1.1)},anchor=south west},
        yticklabel style=red,
        ylabel style=red,
        y axis line style=red,
        ytick style=red
] 
\addplot[mark=+, mark size=\MS, color=red,only marks] table[header=true,x expr=\thisrowno{0},x index=0,y expr=\thisrowno{3}/30.68,y index=3,col sep=space]{./data/p3_partitionInfo-240.txt};
\end{axis}
\begin{semilogyaxis}[
  axis y line*=right,
  axis x line=none,
  xmin = -10,
  xmax = 260,
  ymin=1e-5, ymax=1e2,
        yticklabel style=blue,
        ylabel style=blue,
        y axis line style=blue,
        ytick style=blue
]
\addplot[mark=+, mark size=\MS, color=blue,only marks] table[header=true,x expr=\thisrowno{0},x index=0,y expr=(\thisrowno{3}-30.68)/30.68,y index=3,col sep=space]{./data/p3_partitionInfo-240.txt};
\end{semilogyaxis}
\end{tikzpicture}
\begin{tikzpicture}[scale=0.65]
\begin{axis}[
axis y line*=left,
font=\footnotesize,
grid = minor,
grid=both,
grid style={dotted},
xmin = -500,
xmax = 10500,
ymin = 0,
ymax = 1.25,
xlabel={rank~[-]},
ylabel style={at={(0.04,0.7)}, anchor=east},
legend cell align=left,
legend pos= north east,
legend style={red,at={(0.2,1.1)},anchor=south west},
        yticklabel style=red,
        ylabel style=red,
        y axis line style=red,
        ytick style=red
] 
\addplot[mark=+, mark size=\MS, color=red,only marks] table[header=false,x expr=\thisrowno{0},x index=0,y expr=\thisrowno{3}/7.85,y index=3,col sep=space]{./data/p3_partitionInfo-9216.txt};
\end{axis}
\begin{semilogyaxis}[
  axis y line*=right,
  axis x line=none,
  xmin = -500,
  xmax = 10500,
  ymin=1e-5, ymax=1e2,
  ylabel={$(L_{\text{proc}}-L_{\text{median}})/L_{\text{mean}}$~[-]},
  ylabel style={blue,at={(1.35,0.15)}, anchor=west},
        yticklabel style=blue,
        ylabel style=blue,
        y axis line style=blue,
        ytick style=blue
]
\addplot[mark=+, mark size=\MS, color=blue,only marks] table[header=true,x expr=\thisrowno{0},x index=0,y expr=(\thisrowno{3}-7.89)/7.85,y index=3,col sep=space]{./data/p3_partitionInfo-9216.txt};
\end{semilogyaxis}
\end{tikzpicture}
\caption{Relative measured load (Case B) over MPI ranks (red) and positive deviation from median in 
logarithmic scale (blue), and the number of processes are $240$ (left) and $9216$ (right).}
\label{fig_plume_pload_distribution_p3}
\end{figure}
For illustrating the scaling characteristics in more detail, \figref{fig_plume_pload_distribution_p6} and 
\figref{fig_plume_pload_distribution_p3} depict partition information regarding measured loads of assigned elements for 
the smallest and largest number of MPI ranks chosen for Case A and B respectively. It has to be mentioned that the 
loads are those utilized for the distribution, however they may differ from the actual loads of the subsequent 
calculations.
\figref{fig_plume_pload_distribution_p6} shows that for Case A, the assigned loads of all processes scaled by the mean 
value (red dots) are within a narrow band around unity. The relative deviations in logarithmic scale show that for the 
processes with a load greater than the median (blue dots) the deviations are below one percent. When neglecting the 
single anomalous point for the last MPI ranks, the deviation is on the order of $1\e{-4}$ for the case with 240 
processes. The reason and impact of the increased load of the last process is subject to ongoing 
investigations. All in all, the small deviations of the processes with a larger load than the median, results in an even 
load distribution and small idling times of processes.
In contrast, \figref{fig_plume_pload_distribution_p3} illustrates that for Case B the load is distributed less 
homogeneously. For 240 processes, the relative positive deviation increases to an order of $1\e{-3}$ and 
for 9216 processes to $1\e{-1}$. This significant deviation for a large number of processes leads to an idling 
of the remaining system and, thus, a decreased performance. As exactly those processes deviating from the average load 
band are containing the particles, it is evident that the increased load inhomogeneity is due to the accumulated 
particle time, which is now, at least, in the order of the more homogeneous DG time. The band-width of the load 
deviation could be reduced by shifting elements to larger ranks, where no particles are included. However, this would 
only have a limited impact as those processes constitute only one forth, while the even smaller ratio of 
particle-laden elements are only clustered in isolated sections of the space filling curve.

\section{Summary and Conclusions}\label{sec:conclusions}
In this paper, we have presented a load balancing strategy based on time measurement, where the pure 
computational time per element was determined. The computational effort consists of an approximately 
constant part for the field solver and a dynamic part for the particle treatment. 
To cope with the dynamically changing load, which cannot be determined a priori, the computational time 
was measured in each cell of the domain. This enables an improved load balance over a wide range of applications. 
The new load balance strategy is applied to a plasma expansion scenario~\cite{Copplestone2016}. 
The case demonstrates the load imbalance problem with a constant particle-element weight and an increasing 
particle-element ratio. A significant improvement of the performance can be achieved by pursuing the time 
measurement strategy. Nonetheless, single elements assigned to single MPI processes can make a homogeneous 
load distribution impossible. This is especially encountered for a large number of MPI processes, where a 
large number of particles accumulate in individual elements. A possible solution is a hybrid parallelization 
approach utilizing OpenMP, allowing for larger MPI domains and thus efficiently moving the scaling limit to a 
higher number of CPUs.
In the future, the presented load balance strategy will be applied to coupled PIC-DSMC simulations, which typically exhibit a strong load imbalance.

\section{Acknowledgements}
\label{sec:acknowledgements}
We gratefully acknowledge the Deutsche Forschungsgemeinschaft (DFG) for funding within the projects 
"Kinetic Algorithms for the Maxwell-Boltzmann System and the Simulation of Magnetospheric Propulsion Systems" 
and 
"Coupled PIC-DSMC-Simulation of Laser Driven Ablative Gas Expansions". 
The latter being a sub project of the Collaborative Research Center (SFB) 716 at the University of Stuttgart.
Computational resources have been provided by the Bundes-H\"ochstleistungsrechenzentrum Stuttgart (HLRS).

%
%
%
%
%
\bibliographystyle{plain}
\bibliography{./loc_IMPD.bib}

\end{document}